\documentclass[aps,prl,reprint,superscriptaddres,showpacs]{revtex4-1}

\usepackage{graphicx}
\usepackage{dcolumn}
\usepackage{bm}
\usepackage{times}
\usepackage{color}
\usepackage{braket}

\begin{document}

\bibliographystyle{unsrt}
\preprint{J. Houel {\em et al.} \today}

\title{Probing single charge fluctuations in a semiconductor with laser spectroscopy on a quantum dot}

\author{J. Houel}
\affiliation{Department of Physics, University of Basel, Klingelbergstrasse 82, CH-4056 Basel, Switzerland}

\author{A. Kuhlmann}
\affiliation{Department of Physics, University of Basel, Klingelbergstrasse 82, CH-4056 Basel, Switzerland}

\author{L. Greuter}
\affiliation{Department of Physics, University of Basel, Klingelbergstrasse 82, CH-4056 Basel, Switzerland}

\author{F. Xue}
\affiliation{Department of Physics, University of Basel, Klingelbergstrasse 82, CH-4056 Basel, Switzerland}

\author{M. Poggio}
\affiliation{Department of Physics, University of Basel, Klingelbergstrasse 82, CH-4056 Basel, Switzerland}

\author{B. D. Gerardot}
\affiliation{School of Engineering and Physical Sciences, Heriot-Watt University, Edinburgh EH14 4AS, UK}

\author{P. A. Dalgarno}
\affiliation{School of Engineering and Physical Sciences, Heriot-Watt University, Edinburgh EH14 4AS, UK}

\author{A. Badolato}
\affiliation{Department of Physics and Astronomy, University of Rochester, Rochester, New York 14627, USA}

\author{P. M. Petroff}
\affiliation{Materials Department, University of California, Santa Barbara, California 93106, USA}

\author{A. Ludwig}
\affiliation{Lehrstuhl f\"{u}r Angewandte Festk\"{o}rperphysik, Ruhr-Universit\"{a}t Bochum, D-44780 Bochum, Germany}

\author{D. Reuter}
\affiliation{Lehrstuhl f\"{u}r Angewandte Festk\"{o}rperphysik, Ruhr-Universit\"{a}t Bochum, D-44780 Bochum, Germany}

\author{A. D. Wieck}
\affiliation{Lehrstuhl f\"{u}r Angewandte Festk\"{o}rperphysik, Ruhr-Universit\"{a}t Bochum, D-44780 Bochum, Germany}

\author{R. J. Warburton}
\affiliation{Department of Physics, University of Basel, Klingelbergstrasse 82, CH-4056 Basel, Switzerland}

\date{\today}


%\keywords

\begin{abstract}
We probe local charge fluctuations in a semiconductor via laser spectroscopy on a nearby self-assembled quantum dot. We demonstrate that the quantum dot is sensitive to changes in the local environment at the single charge level. By controlling the charge state of localized defects, we are able to infer the distance of the defects from the quantum dot with $\pm 5$ nm resolution. The results identify and quantify the main source of charge noise in the commonly-used optical field-effect devices. Based on this understanding we achieve routinely close-to-transform-limited quantum dot optical linewidths.
\end{abstract}

\maketitle

Condensed matter systems, notably quantum dots in III-V semiconductors and color centers in diamond, are very attractive as the building blocks for quantum light sources \cite{Shields} and spin qubits \cite{Hanson}. For instance, an InGaAs quantum dot is a robust, high repetition rate, narrow linewidth source of on-demand single photons and polarization-entangled photons, properties not shared by any other emitter. In the future, the demands placed on the quality of the single photons will increase. For instance, the creation of remote entanglement via photon interference and associated applications as a quantum repeater require Fourier-transform-limited single photons, i.e.\ wavepackets with a spectral bandwidth determined only by the radiative lifetime. This is hard to achieve in a semiconductor. On the one hand, a quantum dot is extremely sensitive to the local electric field via the Stark effect \cite{Alen1,Vamivakas1} leading to a stringent limit on the acceptable charge noise. Charge noise can also lead to spin dephasing \cite{Greilich,DeGreve}. On the other hand, phonons in the host semiconductor can lead to dephasing \cite{Ramsay}. However, at low temperature and with weak optical excitation, phonon scattering is suppressed in a quantum dot by the strong quantum confinement \cite{Bayer,Langbein}, and the remaining broadening arises from relatively slow fluctuations of the environment leading to spectral fluctuations \cite{Hogele}. Transform-limited lines have not been routinely achieved, with typical optical linewidths a factor of at least 2 or 3 above the theoretical limit \cite{Hogele,Atature,Xu,Vamivakas2}. While spectral fluctuations in self-assembled quantum dots have been investigated with non-resonant excitation \cite{Robinson,Berthelot}, their origin in the case of true resonant excitation is not known with any precision and are potentially complex with contributions from a number of sources. Furthermore, spectral fluctuations are a common feature in condensed matter systems, arising also in diamond \cite{Robledo}, semiconductor nanocrystals \cite{Muller} and nanowires \cite{Sallen}.

We report new insights into local charge fluctuations in a semiconductor. High resolution laser spectroscopy on a single quantum dot is used as an ultra-sensitive sensor of the local environment. We observe single charge fluctuations in the occupation of a small number of defects located within $\sim 100$ nm of the quantum dot. We control the occupation of these close-by defects with an additional non-resonant excitation. Once the defects are fully occupied, there is a strong suppression of the charge noise. This understanding is tested in a new heterostructure in which the fluctuators are positioned further away from the quantum dot. As predicted by our modelq, this change reduces significantly the quantum dot optical linewidth, making the observation of close-to- transform-limited linewidths routine.

The InGaAs quantum dots are embedded in a Schottky diode \cite{Drexler,Warburton}, Fig.\ 1(a). They are separated from an n$^{+}$ back contact by a $d_{\rm tun}=25$ nm thick GaAs tunnel barrier. Directly on top of the dots is a capping layer of thickness $d_{\rm cap}$, 30 nm in samples A and B, followed by a blocking barrier, an AlAs/GaAs superlattice: $d_{\rm SL}=120$ nm in sample A, 240 nm sample B. Sample C has $d_{\rm cap}=150$ nm and $d_{\rm SL}=240$ nm. Samples B and C were grown under identical conditions. The samples are processed with Ohmic contacts to the back contact, grounded in the experiment, and with a semi-transparent gate electrode on the surface to which a gate voltage, $V_{g}$, is applied. Laser spectroscopy is carried out on the charged exciton $X^{1-}$ at 4.2 K by focusing the linearly-polarized output of a 1 MHz linewidth laser to a $\sim 0.5$ $\mu$m spot on the sample surface. The power of the resonant laser is $\sim 1$ nW to avoid power broadening. The key advance here is to illuminate the sample simultaneously with a weak non-resonant source at 830 nm, Fig.\ 1(b), with power $P$. Resonant excitation of the quantum dot is detected either with differential reflectivity $\Delta R/R$ \cite{Alen2} including a filter to reject the 830 nm light Fig.\ 1(b) or with resonance fluorescence exploiting a dark field technique \cite{Vamivakas2}. The integration time per point is typically 500 (100) ms in $\Delta R/R$ (resonance fluorescence). Spectra are recorded either by sweeping $V_{g}$ (changing the detuning via the Stark effect) or by tuning the laser.

\begin{figure}[t]
\vspace{0.0cm}                                 
\centering                                     
\includegraphics[width=1.0\linewidth]{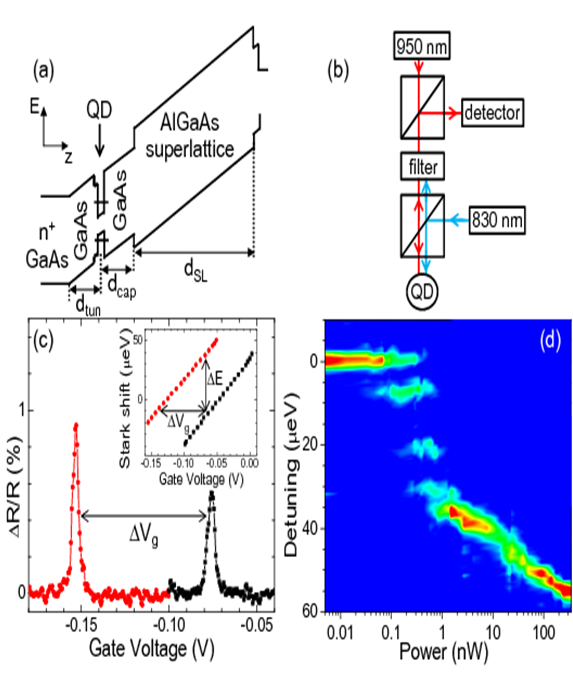}   
\vspace{0.0cm} 
\caption[]{(a) Band diagram of the devices. (b) The optical set-up for $\Delta R/R$ measurements. (c) $\Delta R/R$ versus gate voltage for constant resonant laser wavelength (951.1150 nm) and power (1.0 nW) for a quantum dot in sample A ($d_{\rm cap}=30$ nm) both without (black) and with (red) $P=325$ nW of 830 nm laser light. The inset shows the resonance position versus $V_{g}$. The Stark shift depends linearly on voltage away from the plateau edges; the Stark parameter decreases by only 10\% at $P=325$ nW. (d) Color-scale plot (linear scale, blue: 0.061\%; red: 0.61\%) of $\Delta R/R$ versus non-resonant laser power $P$.}   
\label{figure1}                                       
\end{figure}

Fig.\ 1(c) shows typical laser spectroscopy results both without and with ``high" non-resonant excitation, $P=325$ nW. In both cases, the absorption lines are close to Lorentzians with linewidth 2.5 $\mu$eV. The radiative lifetime at this wavelength is 800 ps \cite{Dalgarno}, implying transform-limited linewidths of $0.8$ $\mu$eV, a factor of 3 smaller than observed in the experiment. Other groups achieve similar linewidths \cite{Hogele,Atature,Xu,Vamivakas2}. The main effect of the non-resonant excitation is to shift the resonance to more negative voltages, in this case by $\Delta V_{g}=-80$ mV, for the same laser wavelength, equivalently a blue-shift of $\Delta E=60$ $\mu$eV for the same gate voltage, Fig.\ 1(c). Fig.\ 1(d) shows $\Delta R/R$ over 4 decades of $P$. Remarkably, the dot evolves from the low-$P$ region (single Lorentzian line independent of $P$) to the high-$P$ region (single Lorentzian line shifting monotonically with $P$) via a series of steps. These steps occur rather abruptly, over just a decade in $P$. For this particular quantum dot, 4 steps (equivalently 5 $\Delta R/R$ lines) are observed. The energy separation of the lines varies from about 4 to 10 $\mu$eV, and the linecuts, Fig.\ 2(c)-(e), show that within each line there is also a sub-structure. The observation of these absorption steps and their behavior as a function of the control parameter $P$ constitute our main experimental discovery.

We find that the $P=0$ and $P=100$ nW behavior are very similar for all dots. Also, the intrinsic properties (radiative lifetime, Stark shift, Coulomb shifts on charging) are all broadly similar. Despite this, the transition region is highly dot dependent. The number of steps lies typically between 3 and 6; the energy separations between the lines lie between $\sim 4$ and 20 $\mu$eV (for sample A) with each quantum dot having its own unique ``finger print" in the $P$-dependence. We therefore look for an explanation of the steps in terms of the dots' environment, i.e.\ a mesoscopic effect. 

Our hypothesis is that nonresonant excitation creates holes at the capping layer/blocking barrier interface, Fig.\ 1(a). 830 nm light creates electron-hole pairs in the wetting layer. The electrons relax rapidly to the back contact, the holes to the capping layer/blocking barrier interface where at low temperature they can be trapped, creating a positive space charge in the device. The trapped holes mean that the same electric field is achieved at the location of the quantum dot only by applying a more negative voltage to the gate, consistent with Fig.\ 1(c). At large $P$, a 2D hole gas is formed, and the shift in $V_{g}$ of the optical resonance allows the hole density $N_{h}$ to be estimated. For intermediate $P$ where we observe the steps, the hole density can be estimated for sample B to be $\sim 10^{10}$ cm$^{-2}$, similar to reported values at the metal-insulator transition \cite{Simmons}. The steps arise in the localization regime. In particular, the steps reflect a change of just one hole in occupation of the localization centers close to the dot. Quantitatively, occupying a localization center immediately above a quantum dot at $d_{\rm cap}=30$ nm changes the electric field by $-1.50$ kVcm$^{-1}$ (taking into account the image charge in the back contact), shifting the optical resonance by 20 $\mu$eV via the Stark shift. This corresponds closely to the maximum observed step separation. This, and the agreement with our simulations (below), justifies our hypothesis. Smaller steps arise from the occupation of localization centers which are laterally displaced.

Our interpretation leads to two immediate results. First, the location of the energy line of the quantum dot is a direct measure of the number of charges stored directly above the quantum dot. In the low-$P$ regime, the quantum dot senses the nearby environment with single charge resolution. Secondly, the number of steps observed equals the number of holes which can be trapped above the dot, 4 for the dot in Fig.\ 1(d).

\begin{figure}[t]
\vspace{0.0cm}                                 
\centering                                     
\includegraphics[width=1.0\linewidth]{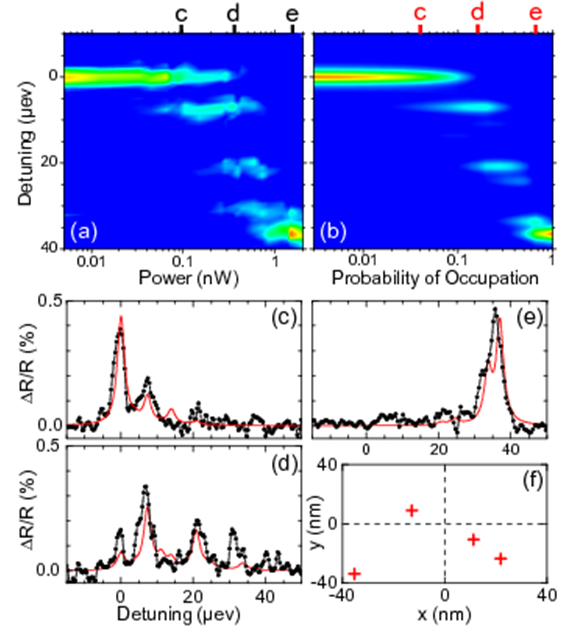}   
\vspace{0.0cm} 
\caption[]{(a) $\Delta R/R$ versus $P$ for a quantum dot in sample A, as in Fig.\ 1(d). (b) Monte Carlo simulation with 4 hole localization centers located above the dot with $r_{i}=(32.0,15.4,15.7,48.8)$ nm, $\alpha_{i}=(5.0,1.5,1.5,1.0)$, $\Gamma=2.5$ $\mu$eV, $\Gamma_{L}=1.0$ $\mu$m and $N=2,500$ (parameters described in the text). (c)-(e) Line cuts at four different values of $P$ showing experimental data (black) and simulation (red). (f) Lateral location of localization centers with dot at $r=0$.}  
\label{figure2}                                       
\end{figure}

We underpin our experimental results with a Monte Carlo simulation of the effects of occupying an array of valence band localization centers at the capping layer/blocking barrier interface. We take an array of localization centers all distance $d_{\rm cap}$ above the quantum dots but at different locations ${\bf r}_{i}=(r_{i},\theta_{i})$ within the 2D plane. We position by hand a small number of localization centers, between 1 and 4, each with $r \le 50$ nm. Additionally, we take a full 2D array of randomly placed defects with 2D density $N_{2D}$. The occupation of a defect changes the local electric field at the quantum dot and hence the absorption spectrum via the Stark effect. This is calculated by, first, calculating the additional electrostatic potential; second, the associated electric field; and third, the energy shift of the exciton via the Stark effect. The Stark shift from the vertical electric field is calculated from the measured Stark effect, i.e.\ from the $V_{g}$-dependence of each particular quantum dot (modeled as a permanent dipole moment in an electric field \cite{Warburton2}). The lateral electric field component cannot be accessed directly in the experiment but the effects are smaller: we assume that there is no linear term (i.e.\ no permanent dipole moment in the lateral plane) and that the quadratic component scales with the fourth power of the wave function extent of the quantum dot ground state which is known reasonably well \cite{Warburton2}. The localization centers $i$ are each occupied with a probability $\alpha_{i} p$ which rises with $p$, the control parameter in the simulation, until it reaches 100\%. $\alpha_{i}$ can vary from center to center and expresses the relative probability of occupying a particular center. With a full 2D array, $\alpha_{i}$ depends on $r_{i}$ through a Gaussian function with full-width-at-half-maximum (FWHM) $\Gamma_{L}$ which describes the spatial extent of the non-resonant beam focus. For the defects directly above the quantum dot, the $\alpha_{i}$ are treated as fit parameters. For a fixed defect distribution and for a given $p$, we occupy the defects with a random number generator; from this charge distribution we calculate the net Stark shift, and at this energy we place a Lorentzian absorption spectrum with FWHM $\Gamma$. The process is repeated $N$ times, keeping the defect distribution constant but each time creating a new charge distribution with the random number generator. The whole procedure is then repeated as a function of $p$. We model the experiment by relating $p$ linearly to the control parameter $P$. 

Our simulation reproduces the steps in the absorption spectra as a function of $P$ for sample A, adding considerable weight to our assertion that the charge fluctuations arise from trapped holes at the capping layer/blocking barrier interface. The exact energy steps turn out to be very sensitive to the locations $r_{i}$ of the localization centers. (The dependence on $\theta_{i}$ is much weaker.) We can match the energies of the steps, their $P$-dependence and the substructure within each step with a set of $r_{i}$ and $N_{2D}=0$. However, we need to depart from $\alpha_{i}=1$ to reproduce the relative intensities of the various lines. Fig.\ 2 shows the result of this procedure: the Monte Carlo simulation, Fig.\ 2(b), reproduces the main experimental features, Fig.\ 2(a). Furthermore, the line-cuts at specific $P$ are in very close agreement with the complicated experimental spectra, Fig.\ 2(c)-(e). The localization center array is shown in Fig.\ 2(f). The different $\alpha_{i}$ presumably reflect some connectivity between the localization centers such that a ``deep" one is much more likely to be occupied than a ``shallow" one. The energy shifts on adding holes to these defects one by one are so sensitive to the set of $r_{i}$ that the random error on each $r_{i}$ is as small as $\pm 5$ nm. In this sense, the experiment provides $\sim \lambda/100$ spatial resolution in the spacings between the localization centers.

We have attempted to reproduce results such as those in Fig.\ 2(a) with just a random distribution of localization centers, $N_{2D}\ne 0$. The large net shift between $P$ ``low" and ``high" pins down $N_{2D}$ to $\sim 10^{10}$ cm$^{-2}$. For this $N_{2D}$, the Monte Carlo simulations predict only in very rare cases $3-5$ steps yet this is the standard experimental result. Furthermore, in the simulation, each line has a strong $P$-dependence, not a feature in the experiment. In the simulations, the only configurations which describe sample A are those with a cluster of localization centers immediately above the dots with otherwise a sparse distribution for $r\le 100$ nm, an extremely unlikely outcome with a random distribution of localization centers. The conclusion is that the localization centers are not randomly distributed in the 2D plane. Instead, the dot itself induces the formation of a small number of localization centers directly above it. The mechanism for this is likely to be the strain field which extends beyond the quantum dot in combination with roughness at the capping layer/blocking layer interface.

\begin{figure}[t]
\vspace{0.0cm}                                 
\centering                                     
\includegraphics[width=1.0\linewidth]{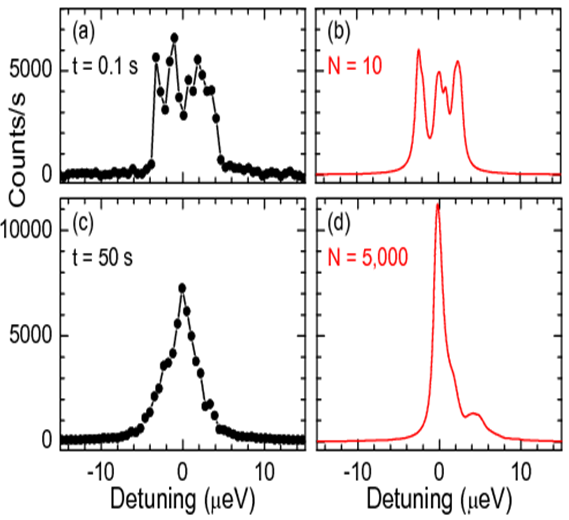}   
\vspace{0.0cm} 
\caption[]{Resonance fluorescence from a single quantum dot in sample B  (1.0 nW at $\lambda=962.2500$ nm) at $P=0$ with integration time per point 0.1 s in (a), 50.0 s in (c). Monte Carlo simulation with $N_{2D}=1.0 \times 10^{10}$ cm$^{-2}$, $\Gamma=0.8$ $\mu$eV, $\Gamma_{L}=10.0$ $\mu$m, $p=4.4$\% (to represent $P=0$) with $N=10$ in (b), $N=5,000$ in (d).}   
\label{figure3}                                       
\end{figure} 

Sample A has $\Gamma=2.5$ $\mu$eV, well above the transform-limit. As described above, this is unlikely to be related to fluctuations in a 2D array of localization centers at the capping layer/blocking barrier interface. The origin of this broadening is not known precisely but there are hints that it is related to the surface of the device. We switch to sample B which clearly demonstrates the consequences of a fluctuating 2D array. Fig.\ 3(a) shows resonance fluorescence from a single dot in sample B. At $P=0$, the FWHM is comparable to those of dots in sample A but there are large fluctuations in the signal which are not reproducible from one spectrum to the next. The fluctuations disappear only when we integrate for more than 50 s per point, Fig.\ 3(c), demonstrating that they have a component at very low frequency (sub-Hz). A characteristic feature is the rather abrupt turn on at negative detunings and the abrupt turn off at positive detunings. Turning on the non-resonant excitation reveals also a series of steps, as in Fig.\ 1(d), and at ``high" P, this sub-Hz frequency component is eliminated. We interpret the $P=0$ results with the Monte Carlo simulations, Fig.\ 3(b), with the hypothesis that the $\mu$eV-scale fluctuations in Fig.\ 3(a) arise from fluctuations amongst a large number of localization centers all with $r \le 100$ nm. With this hypothesis we can reproduce the experiment, Fig.\ 3(b),(d), provided $p$ is small, i.e.\ the defects are each occupied with small probability. The defects (two in this case) directly above the quantum dot are therefore unlikely to be occupied. Only a small fraction of the available configurations are occupied within the measurement time, leading to the changes in spectrum to spectrum. Fig.\ 3(b) reproduces the abrupt turn on/turn off of the spectrum, the FWHM, and the characteristic energy splitting between the sub-peaks using $N_{2D}=1.0 \times 10^{10}$ cm$^{-2}$, $p=4.4$\% and $N=10$. Significantly, the jagged nature of the spectra in Fig.\ 3(a) can only be reproduced with a small homogeneous broadening, $\Gamma=0.8$ $\mu$eV. This is evidence that on short enough time scales, the defect occupation is frozen, and the dot's optical linewidth is close to transform-limited. The behavior for longer integration times, Fig.\ 3(c), is reproduced in the simulations with the same parameters but by increasing $N$, the number of hole configurations, in accordance with the integration time in the experiment, Fig.\ 3(d).  

A key conclusion for sample B is that local fluctuations of hole charges are responsible for the spectral fluctuations and an increase in the optical linewidths in time-integrated spectra above the transform-limit. The first step in reducing the optical linewidths is to take control of these holes. The hole density ($N_{h}=p N_{2D}$) estimated in sample B at $P=0$ is very small, and our experiments have shown that $N_{h}$ is not related to the weak resonant excitation. It is roughly consistent however with the p-type background doping of $\sim 10^{14}$ cm$^{-3}$. Eliminating these holes completely may be challenging. However, based on this new understanding, we have pursued the idea of reducing their influence by forcing the holes to adopt a position further away from the quantum dots. This reduces a hole-induced electric field at the location of the dots by ensuring a closer match between the electric field from an occupied defect and its image charge. Fig.\ 4 shows resonance fluorescence from a dot in sample C with an increased capping layer thickness, $d_{\rm cap}=150$ nm. There are two striking features. First, the linewidth has reduced to 1.4 $\mu$eV. The average linewidth on sample C is 1.60 $\mu$eV with standard deviation 0.22 $\mu$eV. Second, the fluctuations in Fig.\ 3(a) disappear. We attempt to reproduce this behavior in the simulations by keeping $N_{2D}$, $\Gamma$, $N$ and $N_{h}$ exactly the same as for sample B, changing only the capping layer thickness, Fig.\ 4(b). This results in a close-to-Lorentzian line with FWHM 1.1 $\mu$eV, in very close agreement with the experiment. Quantitative understanding of the valence band localization centers has therefore been achieved.

\begin{figure}[t]
\vspace{0.0cm}                                 
\centering                                     
\includegraphics[width=1.0\linewidth]{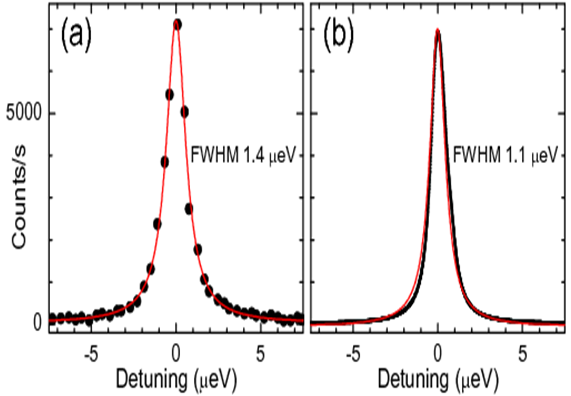}   
\vspace{0.0cm} 
\caption[]{(a) Resonance fluorescence (0.25 nW at $\lambda=951.6040$ nm, $P=0$, 0.1 s integration time) from a dot in Sample C with $d_{\rm cap}=150$ nm (black points; red line Lorentzian fit). (b) Monte Carlo simulation using $N_{2D}=1.0 \times 10^{10}$ cm$^{-2}$, $\Gamma=0.8$ $\mu$eV, $N=10$, $p=4.4$\% (black points; red line Lorentzian fit).}  
\label{figure4}                                       
\end{figure}

We acknowledge financial support from NCCR QSIT, DFG-SPP1285, BMBF QuaHL-Rep 01BQ1035, EPSRC and The Royal Society (BDG) and helpful discussions with Martin Kroner and Alexander H\"{o}gele.

\end{document}